# A Review: Random Walk in Graph Sampling

Xiao Qi

x.qi274@gmail.com

# Abstract

Graph sampling is a technique to pick a subset of vertices and/ or edges from original graph. Among various graph sampling approaches, Traversal Based Sampling (TBS) are widely used due to low cost and feasibility for many cases, in which Simple Random Walk (SRW) and its variants share a large proportion in TBS. We illustrate the foundation SRW and presents the problems of SRW. Based on the problems, we provide a taxonomy of different Random Walk (RW) based graph sampling methods and give an insight to the reason why and how they revise SRW.

our summary includes classical methods, and state-of-art RW-based methods. There are 3 ways to propose new algorithms based on SRW, including SRW and its combinations, modified selection mechanisms, and the graph topology modification. We explained the ideas behind those algorithms, and present detailed pseudo codes. In addition, we add the mathematics behind random walk, and the essence of random walk variants, which is not mentioned in detail in many research papers and literature reviews.

Apart from RW-based methods, SRW also has related with the non-RW and non-TBS methods, we discuss the relationships between SRW and non-RW methods, and the relationships between SRW and non-TBS methods. The relations between these approaches are formally argued and a general framework to bridge theoretical analysis and practical implementation is provided.

# Table of Contents

# 1. Introduction

In recent years, online social networks (OSNs) have become more and more popular, so taking advantage of these platforms to do analysis needed has gain much attention. Examples include road networks (Xie and Levinson, 2007; Duan and Lu, 2014), communication networks (Shimbel, 1953; Gupta, Jain and Vaszkun, 2016), online social and professional networks (Ahn *et al.*, 2007; Bos *et al.*, 2018), citation networks (Portenoy, Hullman and West, 2017; McLaren and Bruner, 2022), collaborative networks (Newman, 2001), biological networks (Charitou, Bryan and Lynn, 2016; Zhang and Itan, 2019), etc. However, some of them are large-scale graphs, or access-constraint datasets. Tackling them and efficiently solving graph mining problems may lead to some problems. To meet this challenge, sampling methods have been developed. We will give some examples and illustrate the necessity of graph sampling techniques.

## 1.1 Motivation for graph sampling

Although OSN plays an important role in the many areas, it has several challenges that affect graph learning, analysis, and visualization. Graph sampling techniques can solve these problems in different aspects.

- **Lack of data**. In network science, researchers need dataset to fit regression models that can predict linkages. However, usually social media platforms have privacy policies, so our access to the whole social networks is limited. In these cases, graph sampling can provide frameworks to preserve the key properties and evolving patterns from the original massive network. Therefore, even with small amount of data, researchers can also investigate the formation patterns of social networks and predict the edges between nodes (Krivitsky and Morris, 2017; Krivitsky, Morris and Bojanowski, 2019).

- **Reduce cost.** When analyzing the social networks, researchers need to download and store the data, and also Using graph sampling can reduce the volume of the graphs/ networks while preserving target properties. In our real-world tasks, the budgets, including the storage and time, are limited. Therefore, reducing the cost by drawing sample properly from the original graphs can promote the efficiency of work.

- **Visualization**. Visualization is an efficient and straightforward way to present the data structure of graphs. For example, Gephi (Jacomy *et al.*, 2014) and NetworkX (Hagberg, Schult and Swart, 2008) are visualization and exploration software for graphs and networks. However, for an extremely large dataset containing agents and links between individuals, the images are not suitable for the screens. Therefore, scientists who need to represent and stress some key users or links can utilize graph sampling methods to capture the structures so that simplify the networks.

## 1.2 An overview of existing methods

Graph sampling techniques have a long history in decades and has attracted much attention recently (Kurant *et al.*, 2011; Xu, Lee and Eun, 2014, 2017; Wang *et al.*, 2015; Chiericetti *et al.*, 2016). Generally speaking, it is a technique that can pick a subset of nodes and/ or edges from the original graph, then get a new subgraph of the original graph with the sampled nodes and/ or edges. To satisfy the needs of research, people need to think how to draw "representative" sample, instead of just a sample. Therefore, researchers come up with many methods to get the representative sample graph from the original graphs. Existing graph sampling methods can be divided into 2 categories according to different way of recruiting nodes and edges. One is the Induced Sampling, which picks nodes/ edges randomly and get the edges connected to those chosen nodes. The other one is based on topology, called link-tracing method, also known as Travel-Based Sampling (TBS) (Hu and Lau, 2013; Coscia and Rossi, 2018).

In TBS, most of the methods are based on Simple Random Walk (SRW), which is the dominant method. In addition, other Travel-Based methods like Snowball Sampling (SBS), Depth First Sampling and Breadth First Sampling are related closely with Random Walk, and many sampling methods belonging to TBS can be recognized as variants of SRW.

As a promising research topic, there are also literature reviews related with graph sampling in recent years (Hu and Lau, 2013; Rozemberczki, Kiss and Sarkar, 2020; Cui *et al.*, 2022). Also, as a classical algorithm, random walk on graphs is well studied (Lovász and Winkler, 1995; Masuda, Porter and Lambiotte, 2017; Xia *et al.*, 2020). However, these papers have a macroscopic perspective and only talk about RW-based methods in TBS or just inventory existing RW-based algorithms, without analyzing and distinguishing strategies for improvement. In this review, we survey important and state-of-art RW-based graph sampling methods, and the mechanisms behind RW. We also explain relations between them and methods of other categories (Section 6.2).

## 1.3 Organization

The reminder of this review is as follows: Section 2 define the notation that will be used in the review; once defined, the notations represent the specific meaning. Section 3, Section 4 and Section 5 inventory Random Walk based sampling methods, from the original idea Simple Random Walk (SRW) to the latest variants. Following every sampling method, the estimation framework is described as graph contains not only the individuals but also the interactions between them. Section 6 discusses the basics of random walk on graphs, and relations between RW-based algorithms and the differences between RW-based methods and methods in other categories, such as Uniform Node Sampling (UNS) and Uniform Edge Sampling (UES). In Section 7, we conclude and give possible research directions in the future.

## 2. Notation

Given an original graph $G = (V, E)$, $V$ is the node set and $E$ is the edge set. Denote the number of elements in $V$ as $N$, i.e., the number of nodes in the original graph. The sample graph, which refers to a sample of the original graph, is denoted as $G' = (V', E')$, where $V' \subseteq V$, and $E' \subseteq E$. Denote the neighbor of a node $v$ as $N(v)$. In RW-based graph sampling methods, the degree of node is a key indicator to decide whether accept the candidate node or not. Here we denote the degree of node $v$ by $d_v$.

In real-world graph sampling applications, there are total budget restrictions. The number of actions tracking an edge is denoted as total budget, which is denoted as $B$ here. Usually, $B \geq |V'|$ as RW-based algorithms are likely to backtrack when exploring the original graph. Thus, the sample graph drawn from the original graph has a fewer number of nodes and edges than the budget.

In the following content, we will inventory random walk and its variants. From the original random walk concept to the latest invented graph sampling methods based on random walk. These methods can be classified into 3 different categories, according to different modification ideas. The first category is simple random walk and its combinations; the second category is modification on selection mechanism; and the last is modifications on structures of network.

## 3. Random Walk and combinations

In Section 3, we introduce the basic Simple Random Walk (SRW), which is the foundation of all RW-based graph sampling methods. We also point out the deficiencies of SRW in graph sampling, i.e., the traits of SRW that lead to low accuracy when drawing sample from graphs and estimating specific properties. Based on these shortcomings, we list several important improvement algorithms. These algorithms are improved by combining and restarting the SRWs without changing the way the SRW selects next-hop node.

## 3.1 Simple Random Walk

The term "Random Walk" was first introduced by Karl Pearson in 1905 (Pearson, 1905), and has been studied for several decades as it has different meaning in different context. Therefore, random walk is still a research focus in the recent years. The best-studied model is the lattices random walk (Révész, 2005), which includes one-dimensional random walk and that in higher dimensions. Random walk on graphs is a variant of the model.

Suppose we have a graph $G = (V, E)$, a simple random walk is a process that one of the neighbors of the current node is chosen uniformly to be the next node. If we specify the number of steps $t$, the $t$-step random walk is a process with random variables $X_1, X_2, \ldots, X_t$ such that $X_i \in V$, and $X_{i+1}$ is a node chosen uniformly at random from the neighbors of $X_i$. In graph sampling, SRW collects all the nodes and edges that are reached in the whole process, thus forms a sample graph, where the node set and edge set are definitely the subset of the node set and edge set of the original graph $G$. Here is the pseudo code of simple random walk.

| **Algorithm 1:** Simple Random Walk (SRW) |
| --- |
| Chose a node $v_0$ from node set $V$ as a seed |
| While $i <$ total budget $B$ |
|     Find the neighbors of current node $v_i$ |
|     Choose a node $u$ uniformly from those neighbors of $v_i$. |
|     $v_{i+1} \leftarrow u$ |
|     $i = i + 1$ |

Random walk has a complete theory as a Markovian process. Markovian process is a class of stochastic process that the next state is only dependent on the current state, while independent with the past history at $0, 1, \ldots, t-1$, thus can be expressed in Markov chain theory as:

$$P(X_{t+1}|X_t) = P(X_{t+1}|X_t, X_{t-1}, \ldots, X_0)$$

We regard all nodes in the node set $V$ as $N$ different states in the sample space $\Omega$, and every edge shares the same probability at time $t$, as the next states are chosen uniformly from the neighbor of current state (node). This sampling mechanism makes the simple random walk obviously Markovian as the probability of choosing next node is only dependent on the current node's out-degree. Specifically, for undirected graphs, it only depends on the degree of current node.

As a Markov chain, the Markov properties of SRW on graphs are also well-studied in the past decades. According to the transition probability of SRW, the stationary distribution is proportional to their node degree, while for edges, the stationary distribution is uniform. Therefore, SRW could be utilized to simulate uniform node sampling when NS is not available or too expensive to conduct in some situations.

The conclusions are right for only connected graphs. For disconnected graphs, it is not right as disconnected components are not reached by pure walking mode. Therefore, researchers also added jumping mode to prevent pure walking mode. Thus, a new variant of SRW, random walk with escaping was proposed.

SRW is the simplest random walk algorithm in the review, and the foundation of all different RW-based algorithms. We can see that all variants are either modifying SRW in selecting neighbor mechanism or filtering the sample collected by SRW.

Sampling procedures can get us samples while estimation framework is also necessary if we need to estimate specific properties of the original graph. For estimation framework, as the stationary distributions are known or able to be calculated when the random walks are designed, the probability of being chosen is known, so is the weight of the sample. A commonly used principle to construct unbiased estimators is employing Horvitz Thompson (HT) estimator (Horvitz and Thompson, 1952) for the function of interest. HT estimator is widely used in many papers (Lee, Xu and Eun, 2012; Li *et al.*, 2015, 2019; Zhou, Zhang and Das, 2015; Zhao *et al.*, 2019).

## 3.2 Random Walk with Restart (RWR)

As single SRW has problems of slow mixing and large deviation, researchers came up with simple combinations to overcome the problems. Pan *et al.* (2004) proposed RWR to calculate the affinity between nodes, and Leskovec and Faloutsos (2006) implemented it in the graph sampling task. Assuming that the starting point of random walk is node $v_0$, RWR sets a non-zero, and fixed probability for the random walker to return to the starting node $v_0$ with probability $p$, which is the key difference between RWR and SRW.

**Algorithm 2:** Random Walk with Restart (RWR)
Chose a node $v_0$ from node set $V$ as a seed
While $i <$ total budget $B$
    Sample a random number $r \sim Binomial(p)$
    **if** r=0
        Find the neighbors of current node $v_i$
        Choose a node $u$ uniformly from those neighbors of $v_i$.
        $v_{i+1} \leftarrow u$
    **if** r=1
        $v_{i+1} \leftarrow v_0$
    $i = i + 1$

RWR can explore the graph more exhaustively while produce a connected subgraph as all sampled nodes are connected to the starting point, and. Later, we will show that RWR is a special case of Random Walk with Escaping (RWE) introduced in Section 4.4. RWE restarts SRW at any node in the graph, while RWR restarts SRW in the specific starting node $v_0$.

### 3.3 Multiple Independent Random Walk (MIRW)

One problem of RW is that it is easy to get stuck in the closely knit community. In sampling area, researchers hope the sample can "saturate" the population so that the estimated outcomes are more closer to the population (Gjoka, Smith and Butts, 2013, 2015). Therefore, if we still need random walker to explore the graph, selecting seeds in different communities is the an ideal solution to make the sample saturate the population graph. Gjoka *et al.* (2010) proposed Multiple Independent Random Walk (MIRW) with different seeds. MIRW firstly chooses $l$ different seeds, and then performs independent random walkers from different starting points (seeds). To decide the ending criteria, we split the total budget $B$ to $l$ random walkers, so every walk has $\frac{B}{l}$ steps.

| **Algorithm 3:** Multiple Independent Random Walk (MIRW) |
| --- |
| /*We have $l$ independent random walks; every walk has the following procedure. */ |
| Chose a node $v_0$ from node set $V$ as a seed |
| Sample edge set $E_s^k \leftarrow \emptyset$, where $k = 1,2,...,l$ |
| Sample node set $V_s^k \leftarrow \emptyset$, where $k = 1,2,...,l$ |
| While $i < \frac{B}{l}$ |
|    Find the neighbors of current node $v_i$ |
|    Chose a node $u$ uniformly from those neighbors of $v_i$. |
|    $E_s^k \leftarrow E_s^k \cup \{(v_i, u)\}$ |
|    $V_s^k \leftarrow V_s^k \cup \{u\}$ |
|    $v_{i+1} \leftarrow u$ |
|    $i = i + 1$ |
| $E_s = E_s^1 \cup E_s^2 \cup E_s^3 ... \cup E_s^l$ |
| $V_s = V_s^1 \cup V_s^2 \cup V_s^3 ... \cup V_s^l$   /*Collect nodes and edges of all $l$ random walks*/ |

However, later MIRW was proved have higher error (Ribeiro and Towsley, 2010) and a new random walk algorithm Multi-Dimensional Random Walk (MDRW) was proposed to improve the accuracy.

MIRW has the same problem with many other RW-based algorithms in real-world applications – usually the whole graph is not available. Therefore, in the first node sampling step, we cannot guarantee that the seeds we choose are exactly located in different communities. The other problem of MIRW is that splitting budget to different random walkers may leads to a situation that single walker cannot walk for long enough time to make the single chain converge. If the convergence is not reached, estimation based on the stationary distribution may lead to higher errors. Compare the single random walker and multiple independent random walkers under the same budget $B$, it is a tradeoff between saturated sample and convergence.

## 3.4 Multi-Dimensional Random Walk (MDRW)

MDRW is called Frontier Sampling when initially came up with. It performs $m$ dependent random walks in the graph by an $m$-dimension vector. In MDRW, every time we sample a new node and edge, the vector $L$ will be updated by the latest sampled node. The dependence is reflected by the probability of i-*th* element in $L$ to be chosen. Note that the $m$-dimension RW on $G$ is equivalent to a one-dimension RW on $G^m$. Here is the pseudo code of MDRW.

| **Algorithm 4:** Multi-Dimensional Random Walk (MDRW) |
|---|
| Choose $m$ nodes as the initial vector $L = (v_1, v_2, ..., v_m)$ |
| Sample edge set $E_s \leftarrow \emptyset$ |
| Sample node set $V_s \leftarrow \emptyset$ |
| **while** $k < b$ |
|     Choose a node $v_i \in L$ with the probability $p_i \propto d_i$  /\*i.e., for $v_i$, the probability of being chosen is proportional to its degree. \*/ |
|     Chose a neighbor of $v_i$ randomly, denote the selected neighbor as $u$ |
|     $E_s \leftarrow E_s \cup \{(v_i, u)\}$ |
|     $V_s \leftarrow V_s \cup \{u\}$ |
|     Replace $v_i$ by $u$ in the vector $L$. |

## 3.5 Re-Weighted Random Walk with (RWRW)

RWRW, which combines SRW with reweighting, is also known as Respondent-Driven Sampling (RDS), initially proposed by Heckathorn (1997). Respondent-Driven Sampling originates from Snowball Sampling (SBS) (Goodman, 1961), recruiting sample by referral-chain. RWRW (RDS) employs Hansen-Hurwitz estimator (Hansen and Hurwitz, 1943) to correct the biased caused by SBS collection. Because SBS can be regarded as a special RW-based sampling method (Section 6.2), in practice, researchers usually use SRW to implement SBS. Therefore, RDS can be regarded as a reweighted SRW; that is why RWRW is also called RDS.

The reason why we mention estimators in RWRW part is that it is not a standalone sampling method. No matter how we get the sample, as long as we can get the sampling distributions of the sample, we can correct the bias. If we regard "estimating graph property" as the final goal, the path can be split into 2 parts: sampling and estimation. For a uniform sample, for example, obtained by MHRW, the estimator can be applied directly while for a biased sample, we need to propose an estimator that can correct the bias. In this goal, if we use MHRW and RWRW in the same dataset, "MHRW + direct estimator" and "RWRW + HH-estimator" are doing the same thing and having similar results.

Till now, we have summarized Simple Random Walk and its parallel forms. One may run a SRW to collect data and reweigh the samples in the estimator (Heckathorn, 1997; Salehi *et al.*, 2011), or runs several random walks on the target graph altogether at one time. Multiple random walk can be performed parallelly or dependently (Gjoka *et al.*, 2010; Ribeiro and Towsley, 2010). Whichever the combination form is, the smallest step of SWR is unchanged – choosing a neighbor of current node uniformly, and accepting it. This selection mechanism only relies on the degree of current node, i.e., it only depends on the information at the current state. That is the definition of Markovian process.

In the following context, we will introduce other two different classes of SRW variants. One class modify the selection mechanism in every step while other modifies the structure of graphs and perform SRW on the graphs, to get the stationary we need. Section 4 introduces the variants that modifies selection mechanism in the micro level while the Section 5 will introduce how network topology affect RW-based algorithms.

## 4. Random walk with modified selection mechanism

In this section, we introduce RW-based sampling methods revising the selection mechanism to either get a stationary distribution closer to uniform, or make sure random walkers escape from the closely knit community, so that it can "propagate" on the graph as quickly as possible.

### 4.1 Employ MCMC techniques: Metropolis Hastings Random Walk

SRW has two main disadvantages in graph sampling. The first one is it tends to recruit nodes with higher degree due to its stationary distribution, so the samples are usually biased towards higher degree node, leading to low accuracies. The second disadvantage is that it might be stuck in closely knit community of the seed node. To overcome these disadvantages, Metropolis Hastings Random Walk (MHRW) and Random walk with Escaping (RWE) are proposed to overcome the aforementioned limitations of SRW, respectively.

Metropolis Hastings random walk is adding a filter in every choosing-neighbor step of SRW. Similar to Metropolis Hastings Random Walk algorithm in real number set, we first chose a neighbor of current node as the candidate node, then we compare the quantity of random number and the degree ratio between current node and candidate node to decide whether accept the candidate node or not. The detailed MHRW pseudo code is below.

**Algorithm 5:** Metropolis Hastings Random Walk (MHRW)

Chose a node $v_0$ from node set $V$ as a seed
While $i < B$
  Find the neighbors of current node $v_i$
  Choose a node $u$ uniformly from those neighbors of $v_i$
  Choose a random number $r \sim U(0,1)$
  **if** $r < \frac{d_u}{d_i}$
    $v_{i+1} \leftarrow u$
  **else**
    $v_{i+1} \leftarrow v_i$
  $i = i + 1$

Theoretically, MH algorithm, as a MCMC technique, can simulate any distribution. However, in graph sampling area, statisticians prefer uniform samples as the variance of uniform distribution is 0. Therefore, when MHRW is employed in previous literatures, the target distribution is usually set as uniform— every node in the original graph has equal probability to be chosen, via random walk. Thus, researchers usually use Uniform Walk when mentioning and/ or employing MHRW (Thompson, 2006; Salehi *et al.*, 2011). It is also easy to see that when we use MHRW, with uniform distribution as the target, it equals to Uniform Node Sampling (UNS), which is too expensive to employ. Therefore, Uniform Walk can be regarded as an inexpensive version of uniform node sampling, although it has waste in other aspects – the existence of acceptance probability rejects parts of candidate nodes.

## 4.2 Unifying Simple Random Walk and Metropolis Hastings Random Walk

Li et al., (2015) proposed Rejection Control Metropolis Hastings (RCMH) framework that can unify SRW and MHRW on graphs. Compared with MHRW, RCMH revises the form of acceptance rate while keeping the selection mechanism in SRW. The pseudo code is below.

**Algorithm 6:** Rejection Control Metropolis Hastings (RCMH)

Chose a node $v_0$ uniformly from node set $V$ as a seed
Sample node set $V_s \leftarrow \emptyset$
Sample edge set $E_s \leftarrow \emptyset$
While $i < B$
  Chose a node $u$ uniformly from $N(v_i)$
  Choose a random number $r \sim U(0,1)$
  **if** $r < \left(\frac{d_u}{d_i}\right)^\alpha$ /*$d_i$ is the degree of node $v_i$*/
    $v_{i+1} \leftarrow u$
    $E_s \leftarrow E_s + (v_i, u)$
  **else**
    $v_{i+1} \leftarrow v_i$
  $i = i + 1$

The essence of RCMH is replacing the acceptance rate $\frac{d_v}{d_u}$ by $\left(\frac{d_v}{d_u}\right)^\alpha$. When $\alpha = 1$, RCMH is equivalent to MHRW, while $\alpha = 0$, $q$ is no larger than 1 definitely as it is a random number from [0,1]. Thus, the candidate node will always be accepted. This is a simple random walk. Therefore, RCMH unifies MHRW and RW by a power function. RW has large deviation problem while MHRW has repeated sample problems. Through adjusting $\alpha$, RCMH can achieve a tradeoff between the large deviation problem of RW and the sample rejection problem of MH.

One remaining thing in this framework: Although The author gave a range, in which the $\alpha$ can make the error relatively low when sampling from real-world datasets, the theoretical analysis of the effect of $\alpha$ is unsolved. Alpha links SRW and MHRW in RCMH. This is an interesting work that can motivate academics to find or propose new frameworks unifying different sampling methods.

## 4.3 Random Walk with Escaping (RWE)

Random walk with escaping (RWE) is a widely used variants of SRW which employs additional jumping mode to an arbitrary random node in the node set. For SRW, random walker chooses a neighbor of current node for sure. But in RWE, random walker chooses a mode with probability $\alpha$ to jump, and $1 - \alpha$ to walk. If the random walker chooses walk, then a neighbor of current node is chosen as in SRW. If the jump mode is chosen, a node is chosen with a probability distribution. Usually when people employ MHRW, which produces uniform samples, use uniform distribution in jump mode to speed up the convergence of the Markov chain.

RWE originally came from PageRank (Bianchini, Gori and Scarselli, 2005) for 2 purpose: 1) make the chain both aperiodic and irreducible; 2) enlarge the spectral gap (Avrachenkov, Ribeiro and Towsley, 2010) so that the chain can converge faster. Escaping is also called "random jump" in many applications.

**Algorithm 7:** Random Walk with Escaping (RWE)
Chose a node $v_0$ uniformly from node set $V$ as a seed
The probability of jumping $\alpha$
While $i < B$
   Randomly choose a number $r \sim U(0,1)$
   **if** r > $\alpha$
    choose a node $u$ uniform from $V$
    $v_{i+1} \leftarrow u$
   **else**
    Find the neighbors of current node $v_i$
    Chose a node $u$ uniformly from $N(v_i)$

$$v_{i+1} \leftarrow u$$
$$i = i + 1$$

However, in our real-world application, RWE still has obvious disadvantages. RW-based algorithms are used to explore unknown graph, it is impossible to reach the whole graph when we use RW or other variants. Therefore, choosing an arbitrary node according to specific distributions is not available, thus RWE's application is not available in many situations. In addition, sometimes RWE cannot prevent the scenario that the random walker jumps into the community where it used to be stuck in, especially when the community in the online social network is huge and closely knit.

### 4.4 Non-Backtracking Random Walk (NBRW)

For the two main problems of SRW mentioned above, being stuck in the same community may leads to repeated samples thus results in high errors in graph sampling. To alleviate this, Lee, Xu and Eun (2012) proposed NBRW, deleting the latest visited edge when choosing a neighbor of current node randomly.

**Algorithm 8:** Non-Backtracking Random Walk (NBRW)
Choose a seed $v_0$ randomly from the node set $V$
Sample node set $V_s \leftarrow \emptyset$
Sample edge set $E_s \leftarrow \emptyset$
$i = 0$
**while** $i < B$
  Choose node $u \in N(v_i) \setminus \{v_{i-1}\}$ uniformly, where $v_{i-1}$ is the latest visited node
  $V_s \leftarrow V_s \cup \{u\}$
  $E_s \leftarrow E_s \cup \{(v_i, u)\}$
  $i = i + 1$

Its stationary distribution on nodes is the same as a SWR on the graphs. Therefore, it overcomes the problem of getting stuck in the same community while keeping the same sampling distribution $p_i \propto d_i$ for nodes and reducing the asymptotic variance.

### 4.5 Circulated Neighbor Random Walk (CNRW)

CNRW (Zhou, Zhang and Das, 2015) is a without-replacement sampling method, deleting all ever visited edges. Different from NBRW that only prevents random walker from visiting the latest visited edge, it remembers all visited edges in the history so that a node cannot be reached by a neighbor twice before the node is isolated. Thus, CNRW needs more memory than NBRW, also needs more memory than any other RW-based methods such as SRW and MHRW. However, deleting visited edges may let connected graph disconnected, so the algorithm restores all edges connecting to the node after them are all deleted. Here is the pseudocode.

**Algorithm 9:** Circulated Neighbor Random Walk (CNRW)

Choose a seed randomly from the node set $V$
Sample node set $V_s \leftarrow \emptyset$
Sample edge set $E_s \leftarrow \emptyset$
**while** $i < B$
  Choose a neighbor $u \in N(v_i)$ uniformly
  Delete edge in the edge set, $E \leftarrow E \setminus \{(v_i, u)\}, G \leftarrow (V, E)$
  **if** $d(v_i) = 0$,
    restore all edges linking $v_i$
  $v_{i+1} \leftarrow u$
  $i = i + 1$

For CNRW, the stationary distribution is the same as SRW, i.e., proportional to node degree. An intuitive explanation of random walk without backtracking can alleviate being stuck in the same community is that, ideally, we would like the random walk path to "propagate" to all parts of graph as quickly as possible, instead of being stuck at a small, strongly connected subgraph like one that includes $u, v$ and $w$ (Zhou, Zhang and Das, 2015).

An extended version of CNRW is Groupby Neighbor Random Walk (GNRW) proposed by Zhou, Zhang and Das (2015) in the same paper. It utilizes a partition function $g$ to partition the $N(u)$ into groups, and escape the visited groups. Like CNRW, it does not alter the stationary distribution of Simple Random Walk - no matter how the grouping strategy is designed. Also, similarly, GNRW guarantees a smaller or equal asymptotic variance, according to the conclusion that the stratification of a random walk's path blocks can affect the asymptotic variance of its estimation (Neal, 2004). Aforementioned NBRW (Lee, Xu and Eun, 2012) also used this conclusion to show that non-backtracking random walk always performs better than SRW.

### 4.6 Common Neighbor Awareness Random Walk (CNARW)

Li *et al.* (2019) proposed a graph sampling method by taking both the number of common neighbors and degrees of current node and candidate node into account. For a candidate node $v$, which is a neighbor of current node $u$, the probability of accepting $v$ as the next state is $\frac{C_{uv}}{\min\{d_u, d_v\}}$, where $C_{uv}$ represent the number of common neighbors of node $u$ and $v$. Candidate node $v$ with a high degree and small number of common neighbors with $u$ makes the walker move from $u$ to $v$ with a higher probability. If $v$ is not accepted, the next node (state) will be selected and assessed by the same procedure until being accepted by the condition, instead of being filled by the current node $u$, which is different from MHRW.

**Algorithm 10:** One walking step of CNARW
Input: current node $u$
Output: next-hop node $v$
1 **do**
2      Select $v$ uniformly at random from $u$'s neighbors;
3      Generate a random number $q \in [0, 1]$;
4      Compute $q_{uv} = 1 - \frac{C_{uv}}{\min\{d_u, d_v\}}$;
5 **while** $(q > q_{uv})$;
6 Return $v$;

An intuitive explanation of CNARW is that the less common neighbor of $u$ and $v$ indicates that they are less likely to be in the same community, thus choosing $v$ might make the walker escape from the community. It overcomes the slow convergence of SRW, as simple random walker tends to be stuck in local loops, due to the high cluster feature. Note that using CNARW will be a bit more demanding than any of the previous algorithms, because not all networks can access their neighbors' information when the node is just a candidate node.

In Section 4, we present 6 different ways to revise selection mechanism. Although they revise the selection mechanism by different ways, the purposes are same. They aim to overcome the deficiency of SRW: by designing weighted walk, they either prevent random walker from getting stuck in a single community or get a uniform distribution. so that the sampled data could be more representative.

## 5. Revising the Graph Structure

In this section, we introduce algorithms that revise target graphs' topology to get the desired distribution or fasten the convergence.

### 5.1 Maximum Degree Random Walk (MD)

MD method add self-loops when a node's degree is less than the maximum degree of the graph and perform random walk on the graph whose all nodes have the same maximum degree.

Let the maximum degree of the graph is $C$, node $i$ with degree $d_i$, which is less than $C$, will add $C - d_i$ self-loops so that all nodes have the same degree. Perform random walk on the processed graph, we can conclude that the stationary distribution is a uniform distribution, as the stationary distribution on a graph is proportional to nodes' degrees. Now that all the nodes have the same degree, the distribution should be uniform.

However, the method also has drawbacks. First, when we use graph sampling strategies, the original graphs are not available, so it is impossible to know the maximum degree. To improve that, researchers suggest an extremely large number $C$, so that $C$ can cover all nodes in the graph and get the uniform distribution we want (Bar-Yossef *et al.*, 2000). The larger $C$ is, the closer to uniform distribution the stationary distribution is. But large $C$ will lead to repeated sample problem. Second, adding self-loops even we can get the maximum degree also leads to a repeated sample problem, i.e., nodes with low degree are more likely to be chosen as they need more self-loops.

## 5.2 Generalized Maximum Degree (GMD)

GMD (Li *et al.*, 2015) process the self-loop slightly different from the MD. In MD mentioned above, they add self-loops till the degree of nodes equals to the number of maximum degrees of the whole graph. In GMD, they restrict the number of self-loops that are added into low-degree nodes. Set a constant $C$, if one node's degree is smaller than $C$, then add $C - d_u$ self-loops on the nodes. For nodes with equivalent or higher degree than $C$, we do not add any self-loop on the current node. Therefore, the stationary distribution of GMD can be expressed as:

$$\pi^{\text{gmd}}(v) = \frac{max\{d_v, C\}}{\sum_{u \in V} max\{d_v, C\}}$$

for any $v$ in $V$. Here is the pseudo code.

**Algorithm 11:** Generalized Maximum Degree (GMD)

1: $u \leftarrow$ initial node;
2: $i = 1$;
3: **while** stopping condition does not meet do
4:     $\xi_i \leftarrow$ Geometric $(d_u / \max\{d_u, C\})$; $S_i \leftarrow u$;
5:     Select a node $v$ uniformly at random from $N(u)$;
6:     $u \leftarrow v$; $i \leftarrow i + 1$;
7: **return** $S$ and $\xi$;

Following the thought that increase the probability of a low-degree node being chosen by adding self-loops on it, GMD adjusted the number of the self-loops by $C$. Following the thought that increase the probability of a low-degree node being chosen by adding self-loops on it, GMD also add self-loops on low-degree nodes, but adjusted the number of the self-loops by $C - d_v$. A large $C$ will lead to repeated sample problems, but a small $C$ will result in large deviation problem, as the smaller $C$ is, the more similar the stationary distribution of GMD is with that of SRW. When $C$ is 0, GMD is the same as SRW; while when $C$ is $d_{max}$, GMD is the same as MD. GMD is a framework that unifies SRW and MD altogether by changing the value of $C$, thus it can balance the tradeoff between the drawbacks of both SRW and MD.

## 5.3 Dual Random Walk (DRW)

The main idea of DRW (Zhang *et al.*, 2020) is use maximum clique structures in communities to represent the whole graph. Thus, it is natural to transfer the graph that we want to draw sample from to a superstructure-based graph. However, constructing superstructure-based graph is costly, especially if the target graph is large or unavailable. Therefore, a hybrid superstructure-based graph based on the sampled nodes not all the nodes in a large graph during DRW's process is constructed to reduce the costs. Furthermore, the existence of both the nodes and the cliques in the hybrid superstructure-based graph is to support the dual residence of the random walker.

**Algorithm 12:** Dual Random Walk (DRW)
Input: Sampling budget $B$
Output: Samples: $u_1, u_2, \ldots, u_B$;
1   $\mu_0 \leftarrow$ Initialize a node randomly from the graph;
2   **for** $i \leftarrow 0$ to $B$ **do**
3      $N \leftarrow$ the number of the nodes in $Nei(u_i)$;
4      $Nei[N] \leftarrow$ the neighbors of $u_i$;
5      $conn[N][N] \leftarrow$ the edge set containing node connection relationship of $Nei(u_i)$, e.g., there is an edge between $\alpha_i$ and $\alpha_j$ ($\alpha_i, \alpha_j \in Nei(u_i)$), $conn[i][j] = 1$;
6      Tlimits $\leftarrow$ 0.025;
7      /* Tlimits is set as the limit of the fraction of the current steps of calculating and sorting the degrees of the nodes to the total steps which are required to find the clique $S(u)$*/;
8      $S(u_i) \leftarrow$ FindMaxClique (conn, $N$, Tlimits);
9      $u_i \leftarrow$ hasVisited;
10     $u_{i+1} \leftarrow$ randomSelect($UneiNode(S(u_i))$);

Algorithms in Section 5 provide a new perspective to revise SRW on graphs. They do not revise the selection mechanism in every step; instead, they still perform a uniform walk in every step. The modification lies in the way of revising topology of the target graphs. One principle to revise topology is that the we do not add new nodes in the graph, so that the number of nodes is same with the original graph.

## 6. Discussion

We discuss the theoretical background behind RW-based graph sampling methods in this section, thus analyzing the common problems of the SRW on graphs. For the purpose of getting small errors, there are 4 factors of RW-based sampling methods that are proved to affect the errors. we also investigated relations between RW-based sampling methods and other categories of methods in graph sampling area.

## 6.1 Basics of Random Walk

What makes random walk so prevalent in graph sampling is that they can be modelled by stochastic process. Especially, SRW, the foundation of RW-based sampling methods, can be modeled as a discrete-time Markov process, whose theoretical analysis is mature in many different areas (Jerrum and Sinclair, 1989; Lovász and Winkler, 1995; Masuda, Porter and Lambiotte, 2017).

For Simple Random Walk on a connected graph, as mentioned before, the next state is only dependent on the current state. More specifically, the probability of the neighbor to be chosen is dependent on the degree of the current node. Therefore, the transition probability can be written as:

$$P(u|v) = \begin{cases} \frac{1}{d_v}, & (v,u) = 1 \\ 0, & (v,u) = 0 \end{cases}$$

Therefore, the stationary distribution $\pi$ of Simple Random Walk satisfies the equation

$$\pi = \pi P$$

Solving the equation and we have the stationary distribution $\pi = \frac{d_v}{\sum d_v} = \frac{d_v}{2|E|}$.

SRW and its variants assumes that the random walk would reach equilibrium, so the authors designing different random walks on graph usually prove the uniqueness of stationary distribution of the random walks, thus give the corresponding stationary distributions (Lee, Xu and Eun, 2012; Li *et al.*, 2015; Zhou, Zhang and Das, 2015; Zhang, 2022).

However, knowing the stationary distributions is not enough. In graph sampling area, the samples are collected when the random walkers start walking on the graphs. Markov chains need some steps to reach the corresponding stationary distributions, which is called "burn-in" in MCMC theory (Guruswami, 2016). During the burn-in, the collected sample is not distributed as the stationary distribution. That is where a part of bias come from. To correct or reduce the bias, researchers need to shorten the burn-in of the chains, i.e., speed up the convergence of random walks.

To discuss SRW on graphs and the Markov chains, three fundamental concepts are necessary: *spectral gap, conductance*, and *mixing time*. According to transition probabilities between all states, one can write the corresponding transition matrix $P$, whose summation of elements in every row is equal to 1. According to Perron Frobenius theory for non-negative matrices (Seneta, 2006) we can conclude that $|\lambda_1| = 1$, and the $|\lambda_i| \leq 1$ for all $2 \leq i \leq n$. So here we denote the eigenvalues of $P$ descending as $1 \geq \lambda_1 > \lambda_2 \geq \cdots \geq \lambda_n \geq -1$. Thus, the spectral gap is defined as $\delta = \lambda_1 - \lambda_2 = 1 - \lambda_2$, i.e., the difference between the largest eigenvalue and the second largest eigenvalue (Avrachenkov, Ribeiro and Towsley, 2010). If a Markov chain's transition matrix has a large spectral gap, then it converges relatively fast, otherwise, the chains converge slowly. Spectral gap plays an important role in bounding mixing time of Markov chain.

Conductance is a concept in Markov chain theory, which measures the tendency that of a Markov chain to move out of a subset of states. It is natural to define conductance of a graph that measures how easy a cut is to get out of in one graph. The conductance of a graph is the minimum conductance among all cuts of the graph.

Intuitively, we can get some insights from aforementioned definitions. When a graph has a high conductance, every part of the graph is not difficult for a random walker to get out of, so the mixing time of the random walker is short, while traveling on graphs with low conductance takes longer time to get steady.

In theoretical analysis, the spectral gap can bound the mixing time, and the conductance of graphs (McNew, 2011). For pure walking mode in an unweighted graph, the transition matrix is equal to the walk matrix (Duncan, 2004) of a graph. Therefore, once the neighbor-choosing strategy is fixed, the spectral gap of the walk is unchangeable. Only if we modify the strategies that a random walker takes, we can design the transition matrix in ways that can make it different from the original adjacent matrix and the walk matrix, so that we can get a transition matrix with a larger spectral gap. Avrachenkov, Ribeiro and Towsley (2010) showed that the second largest eigenvalue of the transition matrix of Markov chain can decreased by introducing random restart and enlarged the spectral gap by restarting the random walk. Later, Gjoka *et al.* (2010) use multiple independent Markov chains, and Ribeiro and Towsley (2010) shortened the mixing time of Markov chains by proposing multiple dependent random walkers. Random jump, an important modification, although had been proposed in previous research, was combined under specific budget restrictions to generate frameworks that unify random walk and random jump (Jin *et al.*, 2011; Xu, Lee and Eun, 2014). Their effort set up frameworks that are able to adjust optimal parameters for different datasets. In addition, selection mechanism modification is utilized in improving the rate of convergence of random walkers (Lee, Xu and Eun, 2012; Zhou, Zhang and Das, 2015; Li *et al.*, 2019). Furthermore, the rate of convergence affects the errors for estimated results. Ribeiro and Towsley (2012) gave the conclusion that the error or estimator is inverse proportional to the spectral gap of RW transition matrix.

From the aspect of getting sample through random walk, if we want to get a representative sample from the population graph, we need to overcome two obstacles which always occur in RW-based graph sampling processes: slow mixing chains and local exploration of the graph. Slow mixing chains leads to large part of biased sample, so in the aspect of collecting sample, speeding up the chains can lower the proportion of biased samples. Local exploration, in the aspect of graph, also leads to samples without enough representativeness. Therefore, we need an exhaustive exploration over the graph via random walker to reduce the errors of the results. One operation can solve these 2 problems. It is speeding up the convergence of the corresponding Markov chain, i.e., design random walk process to increase the spectral gap of the transition matrix. From the classical paper proposing PageRank (Bianchini, Gori and Scarselli, 2005) we can also see that adding the jumping mode in the random walk on graph actually changed the transition matrix, which results in both a faster convergence and prevention from getting trapped in the densely connected component. This kind of problems, eventually can be attributed to the designs of transition matrix. This point of view can even be seen from random walkers on simple graphs. Researchers even can utilize optimization methods to calculate the optimal transition matrices for simple graphs (Boyd, Diaconis and Xiao, 2004; Boyd *et al.*, 2006; Yang, 2015).

## 6.2 Relation with Other Sampling Methods

As mentioned in Section 1, Coscia and Rossi (2018) gave a classification of graph sampling methods in their article, Fig.1. existing methods can be classified into 4 different categories. Induced sampling method, deterministic sampling method, RW-based sampling method and non-RW sampling method. In this section we introduced the similarity and differences between RW-based sampling methods and other 3 different sampling methods, so that one can get to know why RW-based sampling methods play an important role and why sampling methods are classified in this way, so that one can improve them in the future. The classification and relationships are much clearer in Fig 1. We use the dashed line between sampling methods to represent that the two methods can switch to each other by changing parameters or a configuration; double arrow between methods means that they are equivalent in the aspect of sampling probability although the sampling probability is reached by different ways; single arrow represents the improvement.

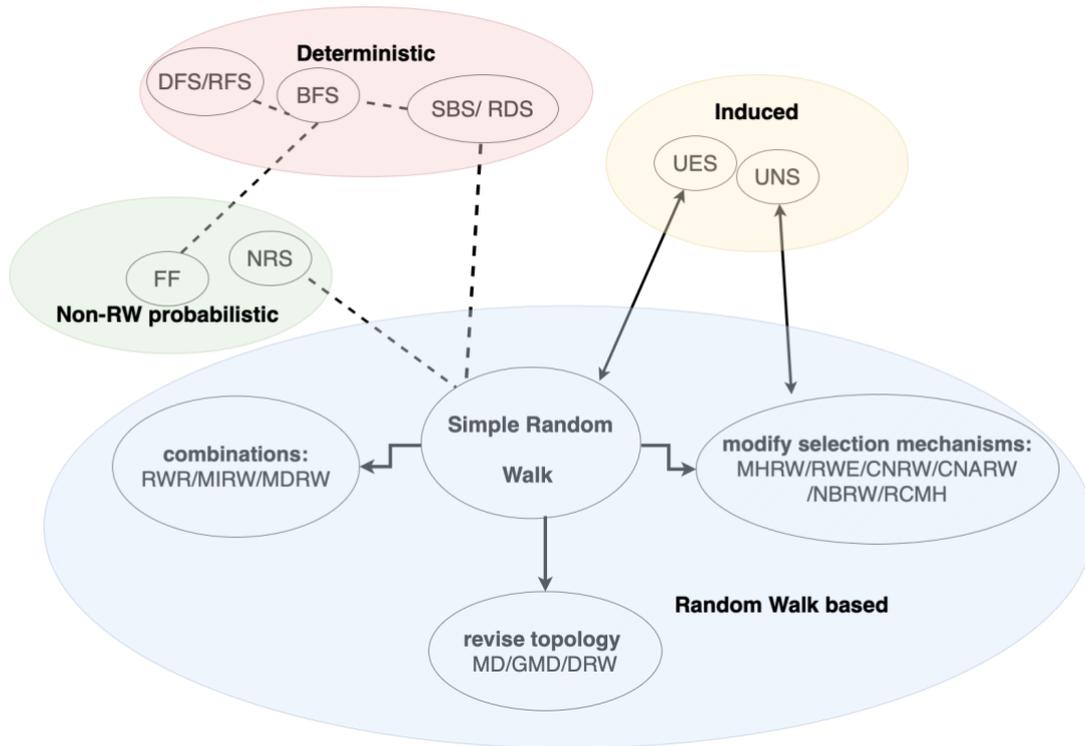

*Figure 1 Map of graph sampling algorithms*

### 6.2.1 Random Walk and Induced Sampling

As mentioned above, the stationary distribution of edges in SRW is uniform, so it is equivalent to the Uniform Edge Sampling (UES). Although the stationary distribution of SRW is not uniform for nodes, later the MHRW is proved to modify the distributions to uniform. That is equivalent to the Uniform Node Sampling (UNS). In fact, Metropolis Hastings algorithm can be used to modify a given distribution to any target distributions. Therefore, we can use Metropolis Hastings Random Walk to simulate Node Sampling with any distribution.

### 6.2.2 Random Walk and Deterministic Algorithms

Deterministic sampling refers to Snowball Sampling (SBS), Depth First Search (DFS), and Breadth First Search (BFS). Doerr and Blenn (2013) unified the DFS, BFS and Random First Search (RFS) into one framework as the following pseudo code shows. They differ in the selection steps. Hu and Lau (2013) summarized the framework in detail. The difference between BFS and DFS is in the implementation of "dequeue ()" methods. For BFS, the first element is selected while for DFS, the last element is selected. BFS is one of the simplest algorithms in graph sampling as it recruits nodes from a starting node, which is called the "root", and every neighbor of the root is explored in the first step. Then, the sample that have already been reached are excluded, and the newly recruited sample nodes' remaining neighbors will be explored. The procedure will continue until the budget is used up. In BFS, all neighbors are recruited either in the last wave or the next wave, therefore the number of sample nodes is unlikely to be the exact number we set in the budget. For example, in a closely connected graph, after 2 waves, the latest wave recruits large number of nodes while the closely connected structure may leads to a larger number of nodes in the next wave. Therefore, without the next wave, the sample size is not reaches, while using the whole sample nodes in the next wave, the budget cannot afford it. That is a limitation of BFS.

**Algorithm 13:** Breadth/ Depth First Search
Initialize the queue with a seed: $V_s \leftarrow \{v_0\}$
$L \leftarrow \emptyset$, $L$ is the set of visited nodes later
Loop until budget B is exhausted:
    Dequeue one element $v = Q.\text{dequeue}()$
    $B \leftarrow B - b, L \leftarrow L + \{v\}$
    For $u \in N(v) \cap u \notin L \cap u \notin Q$, invoke $Q.\text{enqueue}(u)$

Snowball Sampling (Goodman, 1961) can solve this problem as they usually set $k$ neighbors of the current node, instead of all neighbors of current node. Putting the number restriction can control the number of reached nodes, especially when we draw sample from a closely connected graph. If we set the $k = 1$ in snowball sampling, i.e., in every step we only recruit one neighbor the of current node. Besides, deterministic methods do not backtrack while RW-based methods may go back to the previous nodes that ever visited. A non-backtracking SRW is the snowball sampling with $k = 1$.

To sum up, a SBS with $k = 1$ allowing backtracking is equivalent to SRW; if the number of recruited neighbors in every wave is not restricted, the SBS becomes BFS; if the node selection strategy is changed, BFS becomes DFS, or Random Forest Search (RFS) (Doerr and Blenn, 2013; Hu and Lau, 2013). Note that although RFS is not a deterministic sampling method, it can be unified by a framework (Doerr and Blenn, 2013) with BFS and DFS. Also, RWRW, also called RDS, is a combination of SBS and bias-correcting estimation. Since SBS can be implemented by RW, so we usually call it Re-Weighted Random Walk (RWRW).

### 6.2.3 Random Walk and non-Random Walk Methods

Non-RW sampling methods, including Forest Fire (FF) and Neighbor Reviser Sampling (NRS). Although Non-RW and RW are both probabilistic sampling methods, non-RW methods are significantly different from RW.

Forest Fire works as follows: FF performs BFS exploration, i.e., explore every neighbor of the current node. Different from BFS, FF sets a "burning probability" to be a threshold that a node being burnt. Then, FF filters the neighbors that pass the burning probability test and these neighbors are "burnt". Burnt nodes will be explored by BFS in the next step and find their neighbors that pass the burning probability test until the budget is used up.

Usually burning probability let FF choose a part of neighbors. But in extreme situations, FF may choose all of the neighbors or none of any neighbor. The randomness of FF lies in the burning probability test, and all the neighbors are candidate nodes while only neighbors passing the burning probability test can be collected. This concept is similar to the MHRW, but controlling the number is as difficult as BFS since FF never restricts the number of candidate neighbors.

Neighbor Random Sampling (NRS) (Coscia and Rossi, 2018) begins by building a small set of explored nodes using SRW. The majority of NRS's budget goes towards a second phase. The algorithm has already sampled a set of $|V_s|$ nodes. We randomly select a neighbor from the set of $V_s$ nodes that are directly connected to any of the $V$ nodes we have sampled. We also randomly choose a sampled node $v$ from $V_s$. NRS will add $u$ to $V$ if and only if: 1) $G_s$ still has only a single connected component by adding $u$ and removing $v$, and 2) we extract a uniform random number $\alpha < |V_s|/i$, where $i$ is initially set to $V_s$ is set and increases by one for each attempt to add a new neighbor $u$. In practice $|V_s|$ remains constant, but each subsequent attempt to add a new us will be less likely to succeed - because of the increase in $i$. Coscia and Rossi (2018) think NRS is a computationally expensive method because at each potential node addition it has to verify that $G_s$ still has a single connected component (the first condition). Also, given the diminishing acceptance probability, the last units of budget take a long time before they are spent.

For the first condition, it is easy to implement by choosing a sampled node and perform one-step random walk, finding the common neighbor of the candidate node and the chosen sampled node. If at least one common neighbor is in the sample, then we accept the candidate node, otherwise, we find other $u$ and $v$. Similarly, for the second condition, we still need to decide whether accept the node by checking quantities of the random number and $|V_s|/i$.

## 7. Conclusion and Future Work

In this review, we summarize and classify the existing state-of-art RW-based methods used in graph sampling. According to different categories of modifications, they are classified into 3 different categories: Simple Random walk and combinations, Random Walk with revised selection mechanism, and Random Walk with graph structure modification. In every category, we inventory the existing important and state-of-art methods within the category and differentiate them from other categories. We also discuss the spectral theory behind SRW, which is the foundation for the analysis of random walk and designing new RW based algorithms. In addition, we clarify relations between RW-based methods and deterministic methods, the relation between RW-based methods and node/ edge sampling methods.

Random Walk plays a central role in graph sampling. As we have seen in this review, SRW is the core of numerous graph sampling methods, both conceptual and practical. In SRW and the variants of SRW, travelling from one node to another is the basic idea, although some of them decide the next node depending on the information of current node, while some other take last visited node or neighbor information into account. Apart from the RW-based algorithms, in non-RW sampling algorithms, SBS and RDS can be implemented by SRW directly (Hu and Lau, 2013), which also indicates the important role of RW in graph sampling field.

There are several future directions of random walk as there are still space to improve. Usually when researchers use RW-based sampling methods, they derived the corresponding stationary distributions, thus assuming the chain/ stochastic process reaches equilibrium as the algorithms run. This is called burn-in in MCMC theory. Although waiting for the random walker to reach equilibrium has good mathematical properties, long burn-in is wasteful as random walkers on graphs converge slowly (Li *et al.*, 2019). Therefore, one direction is to develop sampling strategies that do not need to wait for convergence of random walkers. Nazi *et al.* (2015) proposed a method that can estimate network properties based on short random walk.

The second possible direction is to add auxiliary information to improve the selection mechanisms. In RW-based sampling methods, most of them use the easy-to-reach information such as degree of current node or the candidate node. In the future work, making use of information as much as possible, including neighbor information and history information, can be a promising research topic.

# Reference:

Ahn, Y.-Y. *et al.* (2007) 'Analysis of topological characteristics of huge online social networking services', in *Proceedings of the 16th international conference on World Wide Web - WWW '07. the 16th international conference*, Banff, Alberta, Canada: ACM Press, p. 835. Available at: https://doi.org/10.1145/1242572.1242685.

Avrachenkov, K., Ribeiro, B. and Towsley, D. (2010) 'Improving Random Walk Estimation Accuracy with Uniform Restarts', in R. Kumar and D. Sivakumar (eds) *Algorithms and Models for the Web-Graph*. Berlin, Heidelberg: Springer Berlin Heidelberg (Lecture Notes in Computer Science), pp. 98–109. Available at: https://doi.org/10.1007/978-3-642-18009-5_10.

Bar-Yossef, Z. *et al.* (2000) 'Approximating Aggregate Queries about Web Pages via Random Walks', in *Proceedings of the 26th International Conference on Very Large Data Bases*. San Francisco, CA, USA: Morgan Kaufmann Publishers Inc. (VLDB '00), pp. 535–544.

Bianchini, M., Gori, M. and Scarselli, F. (2005) 'Inside PageRank'.

Bos, W. van den *et al.* (2018) 'Social network cohesion in school classes promotes prosocial behavior', *PLOS ONE*, 13(4), p. e0194656. Available at: https://doi.org/10.1371/journal.pone.0194656.

Boyd, S. *et al.* (2006) 'Fastest Mixing Markov Chain on a Path', *The American Mathematical Monthly*, 113(1), p. 70. Available at: https://doi.org/10.2307/27641840.

Boyd, S., Diaconis, P. and Xiao, L. (2004) 'Fastest Mixing Markov Chain on a Graph', *SIAM Review*, 46(4), pp. 667–689. Available at: https://doi.org/10.1137/S0036144503423264.

Charitou, T., Bryan, K. and Lynn, D.J. (2016) 'Using biological networks to integrate, visualize and analyze genomics data', *Genetics Selection Evolution*, 48(1), p. 27. Available at: https://doi.org/10.1186/s12711-016-0205-1.

Chiericetti, F. *et al.* (2016) 'On Sampling Nodes in a Network', in *Proceedings of the 25th International Conference on World Wide Web*. WWW '16: 25th International World Wide Web Conference, Montréal Québec Canada: International World Wide Web Conferences Steering Committee, pp. 471–481. Available at: https://doi.org/10.1145/2872427.2883045.

Coscia, M. and Rossi, L. (2018) 'Benchmarking API Costs of Network Sampling Strategies', in *2018 IEEE International Conference on Big Data (Big Data)*. *2018 IEEE International Conference on Big Data (Big Data)*, Seattle, WA, USA: IEEE, pp. 663–672. Available at: https://doi.org/10.1109/BigData.2018.8622486.

Cui, Y. *et al.* (2022) 'A Survey of Sampling Method for Social Media Embeddedness Relationship', *ACM Computing Surveys*, p. 3524105. Available at: https://doi.org/10.1145/3524105.

Doerr, C. and Blenn, N. (2013) 'Metric convergence in social network sampling', in *Proceedings of the 5th ACM workshop on HotPlanet - HotPlanet '13. the 5th ACM workshop*,

Hong Kong, China: ACM Press, p. 45. Available at: https://doi.org/10.1145/2491159.2491168.

Duan, Y. and Lu, F. (2014) 'Robustness of city road networks at different granularities', *Physica A: Statistical Mechanics and its Applications*, 411, pp. 21–34. Available at: https://doi.org/10.1016/j.physa.2014.05.073.

Duncan, A. (2004) 'Powers of the Adjacency Matrix and the Walk Matrix', p. 8.

Gjoka, M. *et al.* (2010) 'Walking in Facebook: A Case Study of Unbiased Sampling of OSNs', in *2010 Proceedings IEEE INFOCOM. IEEE INFOCOM 2010 - IEEE Conference on Computer Communications*, San Diego, CA, USA: IEEE, pp. 1–9. Available at: https://doi.org/10.1109/INFCOM.2010.5462078.

Gjoka, M., Smith, E. and Butts, C.T. (2013) 'Estimating Clique Composition and Size Distributions from Sampled Network Data'. arXiv. Available at: http://arxiv.org/abs/1308.3297 (Accessed: 28 June 2022).

Gjoka, M., Smith, E. and Butts, C.T. (2015) 'Estimating Subgraph Frequencies with or without Attributes from Egocentrically Sampled Data'. arXiv. Available at: http://arxiv.org/abs/1510.08119 (Accessed: 4 July 2022).

Goodman, L.A. (1961) 'Snowball Sampling', *The Annals of Mathematical Statistics*, 32(1), pp. 148–170.

Gupta, L., Jain, R. and Vaszkun, G. (2016) 'Survey of Important Issues in UAV Communication Networks', *IEEE Communications Surveys & Tutorials*, 18(2), pp. 1123–1152. Available at: https://doi.org/10.1109/COMST.2015.2495297.

Guruswami, V. (2016) 'Rapidly Mixing Markov Chains: A Comparison of Techniques (A Survey)'. arXiv. Available at: http://arxiv.org/abs/1603.01512 (Accessed: 17 August 2022).

Hagberg, A.A., Schult, D.A. and Swart, P.J. (2008) 'Exploring Network Structure, Dynamics, and Function using NetworkX', p. 5.

Hansen, M.H. and Hurwitz, W.N. (1943) 'On the Theory of Sampling from Finite Populations', *The Annals of Mathematical Statistics*, 14(4), pp. 333–362. Available at: https://doi.org/10.1214/aoms/1177731356.

Heckathorn, D.D. (1997) 'Respondent-Driven Sampling: A New Approach to the Study of Hidden Populations', *Social Problems*, 44(2), pp. 174–199. Available at: https://doi.org/10.2307/3096941.

Horvitz, D.G. and Thompson, D.J. (1952) 'A Generalization of Sampling Without Replacement from a Finite Universe', *Journal of the American Statistical Association*, 47(260), pp. 663–685. Available at: https://doi.org/10.1080/01621459.1952.10483446.

Hu, P. and Lau, W.C. (2013) 'A Survey and Taxonomy of Graph Sampling'. Available at: https://arxiv.org/abs/1308.5865v1 (Accessed: 2 February 2022).

Jacomy, M. *et al.* (2014) 'ForceAtlas2, a continuous graph layout algorithm for handy network visualization designed for the Gephi software', *PloS one*, 9(6), p. e98679.

Jerrum, M. and Sinclair, A. (1989) 'Approximating the Permanent', p. 30.

Jin, L. *et al.* (2011) 'Albatross Sampling: Robust and Effective Hybrid Vertex Sampling for Social Graphs'.

Krivitsky, P.N. and Morris, M. (2017) 'Inference for social network models from egocentrically sampled data, with application to understanding persistent racial disparities in HIV prevalence in the US', *The Annals of Applied Statistics*, 11(1), pp. 427–455. Available at: https://doi.org/10.1214/16-AOAS1010.

Krivitsky, P.N., Morris, M. and Bojanowski, M. (2019) 'Inference for Exponential-Family Random Graph Models from Egocentrically-Sampled Data with Alter–Alter Relations', p. 23.

Kurant, M. *et al.* (2011) 'Walking on a Graph with a Magnifying Glass: Stratified Sampling via Weighted Random Walks'. arXiv. Available at: http://arxiv.org/abs/1101.5463 (Accessed: 15 August 2022).

Lee, C.-H., Xu, X. and Eun, D.Y. (2012) 'Beyond Random Walk and Metropolis-Hastings Samplers: Why You Should Not Backtrack for Unbiased Graph Sampling'. arXiv. Available at: https://doi.org/10.48550/arXiv.1204.4140.

Leskovec, J. and Faloutsos, C. (2006) 'Sampling from large graphs', in *Proceedings of the 12th ACM SIGKDD international conference on Knowledge discovery and data mining - KDD '06. the 12th ACM SIGKDD international conference*, Philadelphia, PA, USA: ACM Press, p. 631. Available at: https://doi.org/10.1145/1150402.1150479.

Li, R.-H. *et al.* (2015) 'On Random Walk Based Graph Sampling', p. 12.

Li, Y. *et al.* (2019) 'Walking with Perception: Efficient Random Walk Sampling via Common Neighbor Awareness', in *2019 IEEE 35th International Conference on Data Engineering (ICDE). 2019 IEEE 35th International Conference on Data Engineering (ICDE)*, Macao, Macao: IEEE, pp. 962–973. Available at: https://doi.org/10.1109/ICDE.2019.00090.

Lovász, L. and Winkler, P. (1995) 'Mixing of random walks and other diffusions on a graph', in P. Rowlinson (ed.) *Surveys in Combinatorics, 1995*. 1st edn. Cambridge University Press, pp. 119–154. Available at: https://doi.org/10.1017/CBO9780511662096.007.

Masuda, N., Porter, M.A. and Lambiotte, R. (2017) 'Random walks and diffusion on networks', *Physics Reports*, 716–717, pp. 1–58. Available at: https://doi.org/10.1016/j.physrep.2017.07.007.

McLaren, C.D. and Bruner, M.W. (2022) 'Citation network analysis', *International Review of Sport and Exercise Psychology*, 15(1), pp. 179–198. Available at: https://doi.org/10.1080/1750984X.2021.1989705.

McNew, N. (2011) 'The Eigenvalue Gap and Mixing Time'.

Nazi, A. *et al.* (2015) 'Walk, not wait: faster sampling over online social networks', *Proceedings of the VLDB Endowment*, 8(6), pp. 678–689. Available at: https://doi.org/10.14778/2735703.2735707.

Neal, R.M. (2004) *Improving asymptotic variance of MCMC estimators: Non-reversible chains are better*.

Newman, M.E.J. (2001) 'The structure of scientific collaboration networks', *Proceedings of the National Academy of Sciences*, 98(2), pp. 404–409. Available at: https://doi.org/10.1073/pnas.98.2.404.

Pan, J.-Y. *et al.* (2004) 'Automatic multimedia cross-modal correlation discovery', in *Proceedings of the 2004 ACM SIGKDD international conference on Knowledge discovery and data mining - KDD '04. the 2004 ACM SIGKDD international conference*, Seattle, WA, USA: ACM Press, p. 653. Available at: https://doi.org/10.1145/1014052.1014135.

Pearson, K. (1905) 'The Problem of the Random Walk', *Nature*, 72(1865), pp. 294–294. Available at: https://doi.org/10.1038/072294b0.

Portenoy, J., Hullman, J. and West, J.D. (2017) 'Leveraging Citation Networks to Visualize Scholarly Influence Over Time', *Frontiers in Research Metrics and Analytics*, 2, p. 8. Available at: https://doi.org/10.3389/frma.2017.00008.

Révész, P. (2005) *Random Walk in Random and Non-Random Environments*. 2nd edn. WORLD SCIENTIFIC. Available at: https://doi.org/10.1142/5847.

Ribeiro, B. and Towsley, D. (2010) 'Estimating and Sampling Graphs with Multidimensional Random Walks', *arXiv:1002.1751 [cs]* [Preprint]. Available at: http://arxiv.org/abs/1002.1751 (Accessed: 2 March 2022).

Ribeiro, B. and Towsley, D. (2012) 'On the estimation accuracy of degree distributions from graph sampling', in *2012 IEEE 51st IEEE Conference on Decision and Control (CDC). 2012 IEEE 51st Annual Conference on Decision and Control (CDC)*, Maui, HI, USA: IEEE, pp. 5240–5247. Available at: https://doi.org/10.1109/CDC.2012.6425857.

Rozemberczki, B., Kiss, O. and Sarkar, R. (2020) 'Little Ball of Fur: A Python Library for Graph Sampling', in *Proceedings of the 29th ACM International Conference on Information & Knowledge Management. CIKM '20: The 29th ACM International Conference on Information and Knowledge Management*, Virtual Event Ireland: ACM, pp. 3133–3140. Available at: https://doi.org/10.1145/3340531.3412758.

Salehi, M. *et al.* (2011) 'Characterizing Twitter with Respondent-Driven Sampling', in *2011 IEEE Ninth International Conference on Dependable, Autonomic and Secure Computing. 2011 IEEE 9th International Conference on Dependable, Autonomic and Secure Computing (DASC)*, Sydney, Australia: IEEE, pp. 1211–1217. Available at: https://doi.org/10.1109/DASC.2011.196.

Seneta, E. (2006) *Non-negative Matrices and Markov Chains*. Springer Science & Business Media. Available at:

https://books.google.com/books/about/Non_negative_Matrices_and_Markov_Chains.html?id=J3bsjqQBCZUC.

Shimbel, A. (1953) 'Structural parameters of communication networks', *The Bulletin of Mathematical Biophysics*, 15(4), pp. 501–507. Available at: https://doi.org/10.1007/BF02476438.

Thompson, S.K. (2006) 'Targeted Random Walk Designs'. Survey Methodology.

Wang, X. *et al.* (2015) 'Sampling online social networks via heterogeneous statistics', in *2015 IEEE Conference on Computer Communications (INFOCOM)*. IEEE INFOCOM 2015 - IEEE Conference on Computer Communications, Kowloon, Hong Kong: IEEE, pp. 2587–2595. Available at: https://doi.org/10.1109/INFOCOM.2015.7218649.

Xia, F. *et al.* (2020) 'Random Walks: A Review of Algorithms and Applications', *IEEE Transactions on Emerging Topics in Computational Intelligence*, 4(2), pp. 95–107. Available at: https://doi.org/10.1109/TETCI.2019.2952908.

Xie, F. and Levinson, D. (2007) 'Measuring the Structure of Road Networks', *Geographical Analysis*, 39(3), pp. 336–356. Available at: https://doi.org/10.1111/j.1538-4632.2007.00707.x.

Xu, X., Lee, C.-H. and Eun, D.Y. (2014) 'A general framework of hybrid graph sampling for complex network analysis', in *IEEE INFOCOM 2014 - IEEE Conference on Computer Communications*. IEEE INFOCOM 2014 - IEEE Conference on Computer Communications, Toronto, ON, Canada: IEEE, pp. 2795–2803. Available at: https://doi.org/10.1109/INFOCOM.2014.6848229.

Xu, X., Lee, C.-H. and Eun, D.Y. (2017) 'Challenging the limits: Sampling online social networks with cost constraints', in *IEEE INFOCOM 2017 - IEEE Conference on Computer Communications*. IEEE INFOCOM 2017 - IEEE Conference on Computer Communications, Atlanta, GA, USA: IEEE, pp. 1–9. Available at: https://doi.org/10.1109/INFOCOM.2017.8057169.

Yang, Y.K. (2015) 'Numerical Methods for Solving the Fastest Mixing Markov Chain Problem'. Available at: https://www.duo.uio.no/handle/10852/45372 (Accessed: 5 April 2022).

Zhang, L. *et al.* (2020) 'DRaWS: A dual random-walk based sampling method to efficiently estimate distributions of degree and clique size over social networks', *Knowledge-Based Systems*, 198, p. 105891. Available at: https://doi.org/10.1016/j.knosys.2020.105891.

Zhang, L.-C. (2022) 'Graph sampling by lagged random walk', *Stat*, 11(1). Available at: https://doi.org/10.1002/sta4.444.

Zhang, P. and Itan, Y. (2019) 'Biological Network Approaches and Applications in Rare Disease Studies', *Genes*, 10(10), p. 797. Available at: https://doi.org/10.3390/genes10100797.

Zhao, J. *et al.* (2019) 'Sampling online social networks by random walk with indirect jumps', *Data Mining and Knowledge Discovery*, 33(1), pp. 24–57. Available at: https://doi.org/10.1007/s10618-018-0587-5.

Zhou, Z., Zhang, N. and Das, G. (2015) 'Leveraging History for Faster Sampling of Online Social Networks'. arXiv. Available at: https://doi.org/10.48550/arXiv.1505.00079.